\documentclass[a4paper]{jpconf}
\usepackage{graphicx}
\usepackage{subfigure}
\usepackage{wrapfig,rotating}
\begin{document}
\title{Status of the CALICE analog calorimeter technological
  prototypes}

\author{Mark Terwort on behalf of the CALICE collaboration}

\address{DESY, Notkestrasse 85, 22607 Hamburg, Germany}

\ead{mark.terwort@desy.de}

\begin{abstract}
  The CALICE collaboration is currently developing engineering
  prototypes of electromagnetic and hadronic calorimeters for a future
  linear collider detector. This detector is designed to be used in
  particle-flow based event reconstruction. In particular, the
  calorimeters are optimized for the individual reconstruction and
  separation of electromagnetic and hadronic showers. They are
  conceived as sampling calorimeters with tungsten and steel
  absorbers, respectively. Two electromagnetic calorimeters are being
  developed, one with silicon-based active layers and one based on
  scintillator strips that are read out by MPPCs, allowing highly
  granular readout. The analog hadron calorimeter is based on
  scintillating tiles that are also read out individually by silicon
  photomultipliers. The multi-channel, auto-triggered front-end chips
  are integrated into the active layers of the calorimeters and are
  designed for minimal power consumption (power pulsing). The goal of
  the construction of these prototypes is to demonstrate the
  feasibility of building and operating detectors with fully
  integrated front-end electronics. The concept and engineering status
  of these prototypes are reported here.
\end{abstract}

\section{Introduction}

\begin{figure}[t!]
\includegraphics[width=14pc]{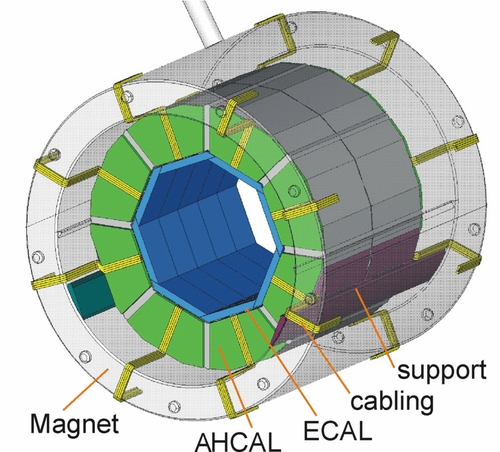}\hspace{2pc}%
\begin{minipage}[b]{14pc}\caption{\label{fig:barrel}Layout of the barrel
    calorimeter system of a LC detector. The ECAL is shown in blue and
    the AHCAL in green, while the structure is surrounded by the
    magnet.}
\end{minipage}
\end{figure}

One of the physics goals at a future linear collider (LC) is the
separation of hadronic W- and Z-boson decays. To achieve this goal a
relative jet energy resolution of $\sim$3\% is necessary, while
typical jet energies are of the order of 100\,GeV and typical single
particle energies are about 10\,GeV. One of the strategies to achieve
such a jet energy resolution is to measure the details of the shower
development in order to separate the showers of individual jet
particles. This information is then combined with the information
obtained from the tracking detectors, such that only the energies of
neutral particles are determined with the calorimeter system. The
approach is known as {\it particle flow}. This implies a number of
challenges for the design and the construction of the calorimeters:
\begin{itemize}
\item High granularity: ECAL (HCAL) cell size $<$ ($\sim$) Moli\`ere
  radius, longitudinally $<$ ($\sim$) 1\,X$_0$,
\item Explosion of channel count: ECAL $\sim 10^8$, HCAL $\sim 10^7$,
\item Compactness: the calorimeters are placed inside the magnet coil,
\item Heat development: power pulsing the integrated front-end
  electronics becomes necessary,
\item 4th dimension: information about the time of each hit with
  respect to the bunch clock is needed.
\end{itemize}
The CALICE collaboration~\cite{CALICE} is developing and testing new
technologies for calorimeters for a future LC experiment.
Technological prototypes are meant to provide answers and solutions to
the challenges listed above. These technological prototypes are
developed in a phase where physics prototypes already provided
valuable information about the physics of hadron showers from
extensive beam test programs~\cite{Marco}. Therefore the main goal is
to demonstrate that the construction of a realistic detector with
fully integrated front-end electronics is feasible.
Figure~\ref{fig:barrel} shows a layout of the calorimeter system of a
LC detector. The ECAL is shown in blue, the analog HCAL option is
shown in green, while the structure is surrounded by the magnet.

Common developments in the collaboration, like a data acquisition
system for all CALICE detectors or readout chips and electronics
boards that are similar for different detectors, allow for a parallel
and cost effective development of different technology options. In the
following, three detectors will be described in more detail. These are
the analog scintillator tile hadron calorimeter (AHCAL) and the SiW
ECAL, as well as the scintillator strip ECAL. For a review of the
physics prototypes of these detectors see~\cite{PPT,SiWECAL,Katsu}.
For a review of the latest status of (semi-) digital hadron
calorimeters see~\cite{DHCAL}.

\section{The AHCAL technological prototype}

\begin{figure}[t!]
\centering
\subfigure[] {\includegraphics[width=2.5in]{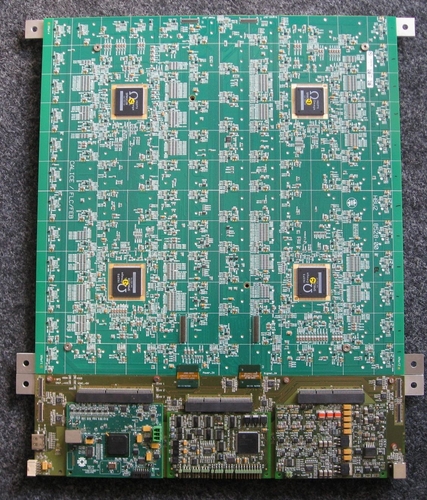}\label{fig:hbu}}
\hspace{1.5cm}
\subfigure[] {\includegraphics[width=2.5in]{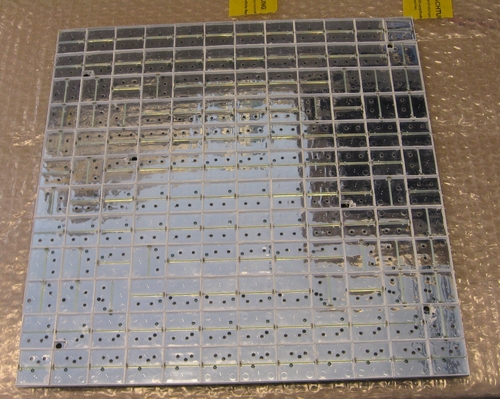}\label{fig:tiles}}
\caption{(a) Photo of the latest basic AHCAL module, equipped with
  four SPIROC2b ASICs. The DAQ interface modules are also shown. (b)
  Photo of tiles that are assembled below an HBU. The orientation of
  the wavelength shifting fibers depends on the position of the holes
  for the pins. This is determined by the details of the PCB design.}
\end{figure}

The CALICE collaboration is currently developing a technological
prototype for an AHCAL~\cite{EPT}. It is based on scintillating
plastic tiles read out by silicon photomultipliers (SiPMs), while
steel or tungsten can be used as absorber material with a thickness of
16\,mm and 10\,mm, respectively. Figure~\ref{fig:barrel} shows how the
AHCAL is placed inside the magnet and outside the ECAL system. It is
divided into four sections along the beam direction, the two endcaps
and two half-barrels. Each half-barrel is further divided into 16
sectors in $\phi$-direction. Each sector consists of 48 layers. The
total thickness is 110\,cm, while the total length of the barrel is
220\,cm. A single active layer is made of three parallel slabs of HBUs
(HCAL base units). Each HBU has a size of $36\times 36$\,cm$^2$ and
has 144 detector channels. The latest HBU version is shown in
Fig.~\ref{fig:hbu}. It is equipped with four SPIROC2b
ASICs~\cite{ASICs} each to read out the photo detectors. On the
backside of the module the scintillating tiles are attached to the PCB
via alignment pins with a nominal distance of 100\,$\mu$m, as shown
in Fig.~\ref{fig:tiles}. Synergies between different CALICE detector
technologies are achieved by deriving the PCBs for an alternative
AHCAL option (by mounting the SiPMs on top of the PCB), as well as for
the scintillator strip ECAL, from the HBU design.

The tiles have a size of $30\times 30\times 3$\,mm$^3$ and are equipped
with a wavelength shifting fiber to guide the light to a SiPM with a
size of 1.26\,mm$^2$ and 796 pixels. The latest batch of SiPMs from
CPTA that are used in the current HBUs have a gain in the range of 1
to 3 million with a noise frequency around 50\,Hz at a 0.5\,MIP
threshold. An alternative option for the tile design is currently
under investigation, where the tiles are read out directly by SiPMs,
e.g. without a wavelength shifting fiber. This reduces the mechanical
complexity, since otherwise the SiPM has to be aligned precisely to
the fiber. First tests already showed that the tiles are very uniform.

The 36-channel SPIROC2b ASIC is specifically designed for LC
operation. It comprises an input DAC for channel-wise bias voltage
adjustment for the SiPMs, it can be power pulsed in order to reach a
power consumption of not more than 25\,$\mu$W per channel (for details
see e.g.~\cite{Peter}) and it can be operated in a self-triggering
mode. Besides setting a global threshold per chip, a channel-wise
tuning of the threshold is possible with a dynamic range of about
0.25\,MIP. A dual-ramp TDC allows for precise time measurements of
individual hits with a resolution of less than 1\,ns~\cite{EPT}.
Strong synergies between the different CALICE calorimeter technologies
are also achieved here, since the SPIROC2b is also used in the
scintillator strip ECAL and a similar chip, called SKIROC, is used in
the SiW ECAL. Detailed laboratory and test beam measurements have lead
to a detailed understanding that allows further developments. For
example, the next generation chip SPIROC2c is currently investigated
in the HBU environment.

\begin{figure}[t!]
\centering
\subfigure[] {\includegraphics[width=2.5in]{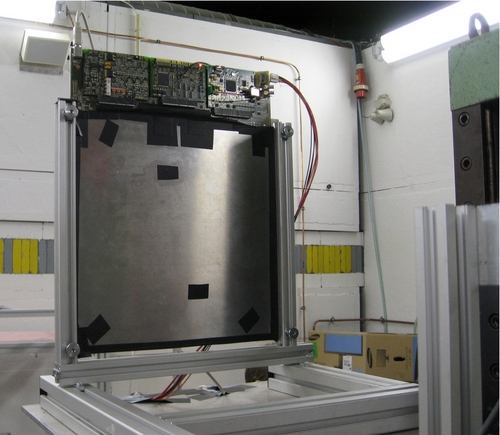}\label{fig:TB}}
\hspace{1.5cm}
\subfigure[] {\includegraphics[width=3.in]{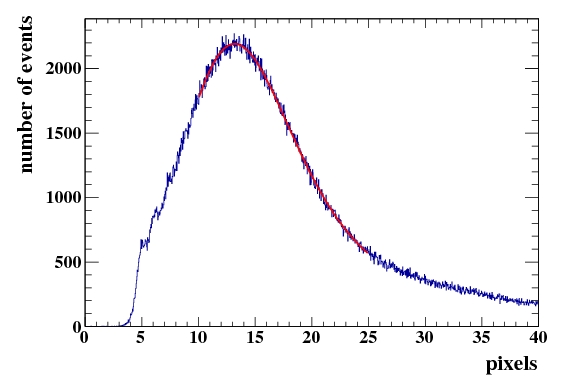}\label{fig:MIP}}
\caption{(a) Setup of the light-tight HBU cassette as it is mounted on
  a movable stage at the DESY test beam facility. (b) Typical MIP
  spectrum obtained from a 2\,GeV electron beam.}
\end{figure}

Four fully equipped new HBUs are currently available and are
extensively tested in the DESY laboratory and test beam facility.
Figure~\ref{fig:TB} shows the setup of a light-tight aluminum HBU cassette
as it is mounted on a movable stage in order to scan all channels with
a 2\,GeV electron beam. The MIP signals are measured in
self-triggering mode and a typical spectrum is shown in
Fig.~\ref{fig:MIP}.

There are several important steps to be done in the near future. Since
there are enough PCBs available to built a full slab, it is possible
now to test the power pulsing mode in a more realistic environment.
The mechanical prototype structure is also in place and therefore
realistic temperature measurements can be performed as well. On the
other hand it is possible now to built a $2\times 2$ HBU layer that
can be used in a hadron beam test. Besides important technical aspects
that can be tested with such a device, it offers the possibility to
measure the radial time distribution of hadron showers when used as an
additional layer behind another CALICE calorimeter.

\section{The ECAL technological prototypes}

The basic requirements for an electromagnetic calorimeter at a
particle flow experiment are similar to the requirements for the
hadron calorimeter. It has to have an extremely high granularity,
while being compact and hermetic. In order to achieve this, tungsten
has been chosen as the absorber material for narrow showers
(X$_0=3.5$\,mm, R$_M=9$\,mm). Two different options for the active
layers are currently under development.

\subsection{SiW ECAL}

\begin{figure}[t!]
\centering
\subfigure[] {\includegraphics[width=2.7in]{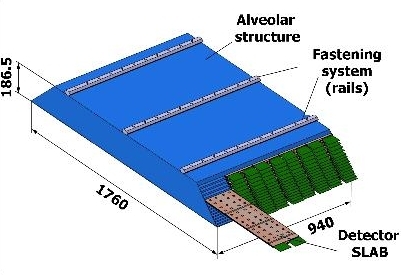}\label{fig:ECAL_structure}}
\hspace{1.5cm}
\subfigure[] {\includegraphics[width=2.2in]{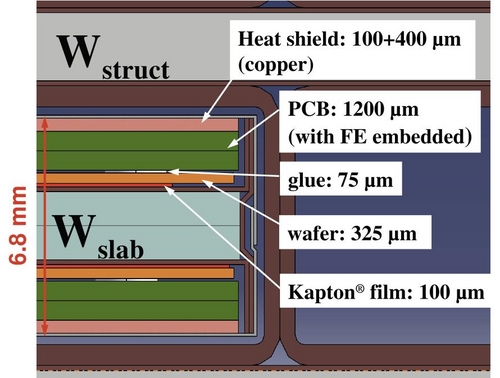}\label{fig:ECAL_structure_2}}
\caption{(a) Schematic of the ECAL alveolar structure, which shows how
  the slabs are inserted into the mechanical housing. (b) Cross
  section of two ECAL layers. The sensitive material and the front-end
  electronics are mounted on both sides of a tungsten carbon composite
  plate.}
\end{figure}

One of the options for the ECAL active layers is to use silicon as
active material. The key parameters of the technological prototype
that is currently developed by the CALICE collaboration
are~\cite{Roman}:
\begin{itemize}
\item Individual cell size of $5.5\times 5.5$\,mm$^2$,
\item Depth of 24\,X$_0$,
\item Thickness of an individual layer of 3.4\,mm and 4.4\,mm
  according to the position within the calorimeter.
\end{itemize}
Figure~\ref{fig:ECAL_structure} shows the geometry of the mechanical
construction that is realized as a tungsten carbon composite to serve
simultaneously as the absorber structure, where individual detector
slabs are inserted. Figure~\ref{fig:ECAL_structure_2} shows a detailed
view of the structure.

\begin{figure}[t!]
\centering
\subfigure[] {\includegraphics[width=3.in]{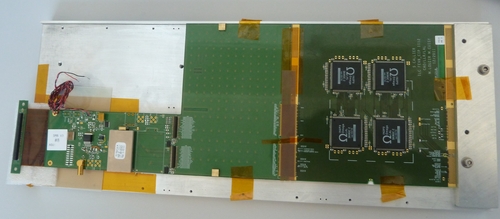}\label{fig:ECAL_layer}}
\hspace{1.5cm}
\subfigure[] {\includegraphics[width=2.in]{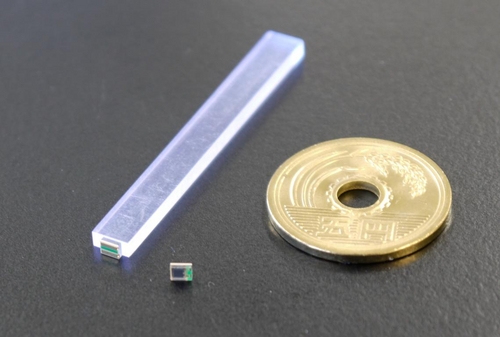}\label{fig:ScECAL_strip}}
\caption{(a) Photo of the FEV8 together with the DAQ interface. (b)
  Scintillator strip coupled to an MPPC as it is used in the active
  layers of the scintillator strip ECAL.}
\end{figure}

Six of the latest modules (FEV8, front-end version) have been tested
recently in the DESY test beam facility. Each FEV8 has been equipped
with four 64-channel SKIROC ASICs, that are specifically designed for
the readout of the ECAL silicon cells. The wafers have a size of
$9\times 9$\,cm$^2$ and comprise 324 pixels.
Figure~\ref{fig:ECAL_layer} shows a photo of the current setup,
including the DAQ interface cards. The main goals for the beam test
are the determination of the signal-to-noise ratio, the establishment
of a calibration procedure for a large number of cells and the
measurement of the homogeneity of the response. During the beam test
it was possible to insert tungsten absorber plates between the active
layers and measure electron showers between 1 and 6\,GeV.
Additionally, this beam test has also successfully demonstated the
functionality of the common CALICE DAQ concept and hardware that ran
smoothly over the complete data taking period. Further beam tests will
follow in the near future, where power pulsing tests and tests with
magnetic field will be performed.

Besides the testing of the existing modules, there are further current
and future challenges. Among these are the optimization of the guard
ring (to suppress cross talk between the modules), the low cost
production of silicon wafers (3000\,m$^2$ for LC detector), the
glueing of the wafers to the PCB, the optimization of the PCB
thickness and the interconnection of individual units (in order to
reduce mechanical stress to the wafers).

\subsection{Scintillator stip ECAL}

A second option for the active ECAL layers is to use scintillator
strips with MPPC readout~\cite{Katsu}. The idea is to use alternating
layers with fine segmentation in x- or z-direction, which leads to an
effective channel size of $5\times 5$\,mm$^2$. After the successful
validation of the concept with a physics prototype, the CALICE
collaboration is currently constructing a technological prototype with
fully integrated front-end electronics. The EBU (ECAL base unit) is
therefore adapted from the AHCAL design and also the SPIROC2b ASIC is
used to read out the MPPCs that are coupled to the strips. Strips are
developed that do not incorporate a wavelength shifting fiber, but are
read out directly with the MPPC, see Fig.~\ref{fig:ScECAL_strip}. The
next important step is to integrate all components and start testing
the complete system in the laboratory and in the DESY test beam
environment.

\section{Summary}

The CALICE collaboration has extensively and successfully tested
physics prototypes for particle flow calorimeters in the past.
Therefore the phase of the validation of particle flow concepts has
finished and lead to the construction and development of technological
prototypes. These prototypes are used to prove the feasibility of the
technological concepts to built realistic linear collider detectors
with fully integrated front-end electronics. Prototypes for
electromagnetic as well as hadronic calorimeters have already taken
data and although the development is still ongoing, basic concepts
could already be addressed and verified:
\begin{itemize}
\item Operation of fully integrated electronics,
\item Power pulsing,
\item Mechanical engineering and
\item Costs and mass production issues.
\end{itemize}
In this report the current status of two electromagnetic and an analog
hadron calorimeter were presented. The tests that have been performed
so far, have lead to an increased understanding of the details of the
systems that allow further developments and larger beam test campaigns
in the near future.

\section*{Acknowledgments}

The author gratefully thanks Karsten Gadow, Erika Garutti, Peter
G\"ottlicher, Oskar Hartbrich, Benjamin Hermberg, Katsushige Kotera,
Roman P\"oschl, Mathias Reinecke, Felix Sefkow and the whole CALICE
collaboration for very useful discussions and valuable contributions
to the results presented here.

% ****************************************************************************
% BIBLIOGRAPHY AREA
% ****************************************************************************

\section*{References}

%\begin{footnotesize}

% ----------------------------------------------------------------------------

%\end{footnotesize}

% ****************************************************************************
% END OF BIBLIOGRAPHY AREA
% ****************************************************************************

\end{document}